\documentclass[aps,pra,twocolumn,groupedaddress]{revtex4}

\usepackage{amsmath,graphicx}
\usepackage{multirow} 

\begin{document}


\title{Ab initio description of nonlinear dynamics of coupled microdisk resonators with application to self-trapping dynamics}
\author{Hamidreza Ramezani$^1$ }
\author{Tsampikos Kottos$^{1,3}$}
\author{V. Shuvayev$^2$}
\author{L. Deych$^2$}
\affiliation{$^1$Department of Physics, Wesleyan University, Middletown, Connecticut 06459, USA}
\affiliation{$^2$Department of Physics, Queens College of the City University of New York (CUNY) Flushing, NY 11367}
\affiliation{$^3$Max-Planck-Institut f\" ur Dynamik und Selbstorganisation, G\" ottingen, Germany}

\date{\today}

\begin{abstract}
{\it Ab initio} approach is used to describe the time evolution of the amplitudes
of whispering gallery modes in a system of coupled microdisk resonators with Kerr nonlinearity. It is shown that this system demonstrates a transition
between Josephson-like nonlinear oscillations and self-trapping behavior. Manifestation of this transition in the dynamics of radiative losses is studied.
\end{abstract}


\maketitle
\section{Introduction}
Whispering gallery modes (WGM) are optical excitations that exist in axially symmetrical optical resonators. In the
geometric optic framework they can be described as waves propagating at large incidence angles along the surface of
the resonator and trapped in an optical potential arising due to total internal reflection. Due to high Q-factors and small mode volumes of WGMs, one should expect a significant enhancement of nonlinear effects in such structures~\cite{BraginskyPhLetA1989,Lin:1994,Braunstein:1996}. Indeed, a variety of nonlinear phenomena such as ultra-low threshold Raman lasing~\cite{VahalaNature2002}, parametrical optical oscillations\cite{Savchenkov:2004},
and optical comb generation~\cite{Strekalov:2009} have been experimentally demonstrated in WGM resonators. Theoretical understanding of these effects, however, relied mostly either on purely phenomenological approaches similar to that of Ref.~\cite{Matsko:2005}, or employed approximations of the mean-field type~\cite{Fomin:2005,Braunstein:1996}.

When several WGM resonators are placed in the proximity of one another, they become optically coupled due to overlapping evanescent tails of their respective modes. The resulting collective optical excitations in linear regime have been observed experimentally in coupled microspheres and micro-disks~
\cite{MoellerOL2005,MukaiyamaPRL2005,Schmidt:2009} and described theoretically within the framework of \emph{ab initio}
approaches~\cite{MiyazakiPRB2000,BoriskinaReview2006,DeychRoslyakPRE2006,Schmidt:2009}. At the same time, the nonlinear optics of coupled microresonators is still in its infancy. While lately there was a significant uptick in theoretical attention to this field~\cite{Irish:2008,Egorov:08,SGHBT10,FACTG10}, no experiments in this area have yet been reported. This rise of interest to optically coupled resonators is fueled by expectations that these systems would allow for a convenient experimental studies by optical means of some important nonlinear effects of generic nature. In this work we demonstrate theoretically that coupled microdisk resonators with instantaneous Kerr nonlinearity is a convenient system, in which such nonlinear effects as Josephson oscillations and self-trapping can be demonstrated. This system is significantly simpler for experimental realization than those considered in Ref.~\cite{Irish:2008,SGHBT10,FACTG10}, and we expect, therefore, that this work will motivate experimental studies of nonlinear optical properties of microdisks and other coupled WGM resonators.
 
Unlike previous theoretical works relying on phenomenological models~\cite{Irish:2008,Egorov:08,SGHBT10,FACTG10}, we use an \emph{ab initio} approach to derive equations describing nonlinear dynamics of collective WGM excitations in this system. We show that under certain conditions these equations can be presented in the form of modified discrete nonlinear Schr\"{o}dinger equation (DNLSE). DNLSE is a prominent model known to describe successfully the dynamics of systems as diverse as Bose-Einstein Condensates
(BEC) in optical lattices, polarons, optical fibers, and biological molecules~\cite{CFK04,LSCASS08,HT99,SGHBT10,FACTG10,
SCWS08,LPWAD10,AGFHCO05,FW98,BKR03}. DNLSE describes such nonlinear phenomena as self-trapping, discrete breathers, etc.
and demonstrate anomalously slow relaxation dynamics~\cite{TA96} (due to boundary losses) similar to the one appeared
in glass materials. This relaxation dynamics has generated recently a great deal of theoretical interest in the context of cold atoms~\cite{KW04,LFO06,GKN08,NHFKG09} and coupled cavities\cite{SGHBT10,FACTG10}. At the same time, while self-trapping in systems with conserving number of particles or energy has been observed experimentally~\cite{AGFHCO05}, it is still yet to be observed in in any physical realization of the DNLSE with losses.

In this work, we demonstrate that the system of coupled microdisk cavities provides a convenient platform for studying nonlinear dynamics in the presence of relaxation. While radiative losses in such a structure are intrinsic to each disk, one can introduce an imbalance of the losses required for anomalous relaxation by coupling one of the disks  to a tapered fiber. Such an arrangement, which is usually used to excite WGMs, is known
to reduce the Q-factor of the fiber-coupled resonator~\cite{PainterPRL2003} providing thereby a faster relaxation
channel through the boundary of the structure. We demonstrate that at certain threshold values of initial
conditions/nonlinearity the dynamics of the system exhibits a transition between nonlinear Josephson-like oscillations
of light intensity between the disks and a self-trapping behavior. We also show that the transition between these
two regimes is manifested in different kinetics of radiative losses in Josephson and self-trapping regimes.
\section{Ab initio dynamical equations}
While we focus in this paper on a nonlinear dynamic of two laterally coupled microdisk resonators, the general dynamic equations can be derived for a more general case of $N$ disks, provided they all lie in the same plane, which is parallel to their surfaces.  In the spectral region of high-Q WGM resonances electromagnetic field of a single disk can be described by a simplified two-dimensional model~\cite{BorselliOE2005}, in which its electric (for TM modes) or magnetic (for TE modes) field is perpendicular to disk's surface and is, therefore, characterized by a single function $F^{(i)}({\bf r},t)$. In this approximation, disks' refractive index is replaced by an effective parameter,
$n_d$, determined from a self-consistency condition \cite{BorselliOE2005}. Two-dimensional approximation was also used to describe electromagnetic field of the planar system of several disks (optical molecules) using coupled integral equations\cite{BoriskinaOL206,BoriskinaOptExpr:2007} as well as modal expansion approach~\cite{Schmidt:2009}. While the formulation in terms of integral equations is convenient for numerical simulations of planar structures, especially with non-circular elements, the modal expansion is more appropriate for developing \emph{ab initio} theoretical description of nonlinear effects in WGM resonators.
\subsection{Single disk in  presence of external incident field and internal polarization}
Before developing the theory of the nonlinear dynamics of WGMs in multiple disks, one has to consider the case of a single disk. We will assume that the nonlinearity does not couple TM and TE polarizations  and limit our consideration to the TM case. This assumption is valid for disks made of materials with small optical anisotropy and in the absence wave mixing processes. In the spectral domain two-dimensional wave equation for the Fourier transformed electric field can be written as
\begin{equation}\label{eq:Maxwell}
\begin{split}
    &\nabla^2 F(r,\phi,\omega)+\frac{n_d^2\omega^2}{c^2}F(r,\phi,\omega)=-\frac{4\pi\omega^2}{c^2}P, \hskip 1.5pt r<R\\
    &\nabla^2 F(r,\phi,\omega)+\frac{\omega^2}{c^2}F(r,\phi,\omega)=0, \hskip 1.5pt r>R
\end{split}
\end{equation}
where $r$, and $\phi$ are, respectively, radial and angular polar coordinates, defined in a coordinate system with the origin at the disk's center, $\omega$ is a spectral parameter, not yet identified with any physical frequency, $n_d$ is the effective refractive index of the disk,  $c$ is vacuum speed of light, and $P$ is the nonlinear polarization. Usually, Eq.\ref{eq:Maxwell} is solved in the context of the scattering problem, when the boundary conditions are of inhomogeneous type  defined in the presence of an external "incident" field. At the same time, nonlinear effects are usually described in terms of dynamics of "modal amplitudes", with modes being solutions of a linear problem with homogeneous (no incident field) boundary conditions. The difficulty for the nonlinear theory of WGMs, which are usually excited by an "incident" field, is the necessity to reconcile these two different types of problem.  In order to achieve this we, following Ref.~\cite{mors53}, introduce a Green's function defined by equations
\begin{equation}\label{eq:Green}
\begin{split}
    &\nabla^2 G(\mathbf{r}-\mathbf{r}^\prime)+\frac{n_d^2\omega^2}{c^2}G(\mathbf{r}-\mathbf{r}^\prime)=-\delta(\mathbf{r}-\mathbf{r}^\prime), \hskip 1.5pt r<R\\
    &\nabla^2 G(\mathbf{r}-\mathbf{r}^\prime)+\frac{\omega^2}{c^2}G(\mathbf{r}-\mathbf{r}^\prime)=-\delta(\mathbf{r}-\mathbf{r}^\prime), \hskip 1.5pt r>R
\end{split}
\end{equation}
and boundary conditions
\begin{equation}\label{eq:bound_cond}
\begin{split}
&G(R-0,r^\prime)=G(R+0,r^\prime)\\
&\left.\frac{\partial G(r,r^\prime)}{\partial r}\right|_{r=R-0}=\left.\frac{\partial G(r,r^\prime)}{\partial r}\right|_{r=R+0}\\
&\left.\frac{\partial\ln{\left[r^{1/2} G(r,r^\prime)\right]}}{\partial r}\right|_{r\rightarrow\infty}=i\frac{\omega}{c},
\end{split}
\end{equation}
where the last of these expressions establishes outgoing field condition at infinity. Using Green's theorem~\cite{mors53} one can express the field in the interior of the disk in terms of the Green's function defined in Eq.~\ref{eq:Green} and boundary values of the field and its derivative:
\begin{equation}\label{eq:field_integral}
\begin{split}
&F_m(r)=\frac{8\pi^2\omega^2}{c^2}\int_0^RG_m(r^\prime,r)P_m(r^\prime)r^\prime dr^\prime \\
&- 2\pi R\left[F_m(R)\frac{\partial G_m(r^\prime,r)}{\partial r^\prime}-G_m(R,r)\frac{\partial F_m(r^\prime)}{\partial r^\prime}\right]_{r^\prime=R}
\end{split}
\end{equation}
where all quantities with subindex $m$ are angular Fourier coefficients of their respective functions: $<\cdot>_m=(1/2\pi)\int_0^{2\pi}<\cdot>(\phi)\exp{[im\phi]}d\phi$. The values of the field and its derivative at the boundary of the disk, $r=R$, are defined via Maxwell boundary conditions demanding continuity of $F$ and its derivative with respect to the radial variable.  In the presence of the incident field, $F_{inc}(r,\phi)$, the field outside of the disk is $F_{inc}(r,\phi)+F_{sc}(r,\phi)$, where $F_{sc}(r,\phi)$ is the field scattered by the disk. Taking into account standard conditions for the incident field to be finite at $r=0$ and for the scattered field to be outgoing at infinity, these fields can be presented as linear combinations of the first order Bessel ($J_m$) and Hankel ($H_m$) functions:
\begin{eqnarray}
F_{inc}&=&\sum_m a_mJ_m(kr)\exp{(im\phi)}\label{eq:inc}\\
F_{sc}&=&\sum_m b_mH_m(kr)\exp{(im\phi)}\label{eq:sc}
\end{eqnarray}
where $k=\omega/c$. Combining Eq.~\ref{eq:field_integral} with the boundary conditions as well as with Eq.~\ref{eq:inc} and \ref{eq:sc} one arrives at the following expression for the internal ($r<R$) field of the disk:
\begin{equation}\label{eq:F_final}
\begin{split}
F_m(\rho)&=8\pi^2x^2\int_0^1G_m(\rho^\prime,\rho)P_m(\rho^\prime)\rho^\prime d\rho^\prime \\
&-\frac{2i}{\pi x\xi_m(x)}a_mJ_m(n_dx\rho)
\end{split}
\end{equation}
where we introduced dimensionless spectral parameter $x=kR$ and radial coordinate $\rho=r/R$. Function $\xi_m(x)$ in the second term of Eq.~\ref{eq:F_final} is defined as
\begin{equation}\label{eq:xi_r}
\xi_m(x)=n_dH_m(x)J_m^\prime(n_dx)-J_m(n_dx)H_m^\prime(x)
\end{equation}
where $f^\prime(z)\equiv df/dz$.

In the absence of the polarization source, Eq.~\ref{eq:F_final} reproduces a standard solution for the internal field in the linear single disk scattering problem. The scattered field in this case is also found in a standard form
\begin{equation}\label{eq:b}
b_m=\alpha_m a_m
\end{equation}
where $\alpha_m$ is a linear scattering amplitude defined as
\begin{equation}\label{eq:alpha}
\alpha_m=\frac{p_m(x)}{\xi_m(x)}
\end{equation}
where $p_m(x)$ is
\begin{equation}\label{eq:p}
p_m(x)=J_m^\prime(x)J_m(n_dx)-n_dJ_m^\prime(n_dx)J_m(x).
\end{equation}
The poles of $\alpha_m(x)$, which we will designate as $x_{m,s}^{(r)}-i\gamma_{m,s}^{(r)}$, are given by complex-valued solutions of equation $\xi(x)=0$ and define spectral positions, $x_{m,s}^{(r)}$, and widths, $\gamma_{m,s}^{(r)}$, of linear WGM resonances in a single disk.   The sub-index $s$ here distinguishes resonances with different radial distributions of the respective internal fields: the value of this index determines the number of maxima of the field in the radial direction. The scattering amplitude has an important property: $\alpha_m(x_{m,s}^{(r)})\equiv -1$, which results in the following single pole approximation valid in the vicinity of a selected  resonance:
\begin{equation}\label{eq:alpha_approx}
\alpha_m=-\frac{i\gamma_{m,s}^{(r)}}{x-x_{m,s}^{(r)}+i\gamma_{m,s}^{(r)}}
\end{equation}

Eq.~\ref{eq:b} and \ref{eq:alpha} solve the linear scattering problem and describe resonant response of the disk to the external excitation. However, complex frequencies $x_{m,s}^{(r)}$ are not eigenfrequencies of the disk, and their respective field distributions are not its eigenfunctions. In order to define true normal modes of the system together with their eigenvalues, one has to solve the wave equation with boundary conditions formulated in the absence of the incident field (homogeneous boundary conditions). The open nature of the optical resonators makes this problem nontrivial. Several different approaches have been developed to introduce a system of functions, generalizing the concept of modes, that could be used as a basis to describe open systems (see, for instance, recent review~\cite{Zaitsev:2010}). For our goals, the most convenient is the approach based on so called Constant Flux Modes, re-introduced in the optical context in Ref.~\cite{ture06} and used recently in Ref.~\cite{Prasad:2010}. These functions are defined as solutions of the following equations
\begin{equation}\label{eq:eigen}
\begin{split}
    &\nabla^2 \psi_{m,s}(\rho,\phi)+n_d^2x^2_{m,s}\psi_{m,s}(r,\phi)=0 \hskip 1.5pt r<R\\
    &\nabla^2 \psi_{m,s}(r,\phi)+x^2\psi_{m,s}(r,\phi)=0, \hskip 1.5pt r>R
\end{split}
\end{equation}
with the same boundary conditions as those introduced in Eq.~\ref{eq:bound_cond} for the Green's function. It is crucial that outside of the disk these functions obey the wave equation with generic spectral parameter $x$ rather than with the eigenvalue $x_{m,s}$.  Only functions defined this way form, in combination with their adjoint counterparts, a complete bi-orthogonal set with an inner product and norm defined over the \emph{interior} of the disk:
\begin{equation}\label{eq:biorth}
\int_0^1\psi_{m,s}(\rho)\left[\overline{\psi}_{l,p}(\rho)\right]^*\rho d\rho=N_{m,s}^2\delta_{m,l}\delta{p,s}
\end{equation}
Adjoint functions $\overline{\psi}_{l,p}(\rho)$ appearing in Eq.~\ref{eq:biorth} are solutions of equations
\begin{equation}\label{eq:adjoint}
\begin{split}
    &\nabla^2 \overline{\psi}_{m,s}(\rho,\phi)+n_d^2[x^2_{m,s}]^*\overline{\psi}_{m,s}(r,\phi)=0 \hskip 1.5pt r<R\\
    &\nabla^2 \overline{\psi}_{m,s}(r,\phi)+x^2\overline{\psi}_{m,s}(r,\phi)=0, \hskip 1.5pt r>R
\end{split}
\end{equation}
with incoming boundary conditions at infinity. The  eigenvalues $x_{m,s}$ are defined by equation
\begin{equation}\label{eq:eigenvalues}
n_dx_{m,s}H_m(x)J_m^\prime(n_dx_{m,s})-xJ_m(n_dx_{m,s})H_m^\prime(x)=0
\end{equation}
and are complex-valued functions of the spectral parameter $x$, which becomes an integration variable upon transformation of the fields back to the time-domain.

Since eigenfunctions $\psi_{m,s}$ obey the same boundary conditions as the Green's function, $G_m(\rho,\rho^\prime)$, they can be used to generate its spectral expansion
\begin{equation}\label{eq:decomp}
G_m(\rho,\rho^\prime)=-\frac{R^2}{2\pi n_d^2}\sum_s\frac{\psi_{m,s}(x_{m,s}n_d\rho)\overline{\psi}^*_{m,s}(x_{m,s}n_d\rho^\prime)}{x^2-x_{m,s}^2}.
\end{equation}
The same set of functions can be used to present the internal field of the disk
\begin{equation}\label{eq:int_field}
F_m(\rho,x)=\sum_sD_{m,s}(x)\psi_{m,s}(x_{m,s}n_d\rho)
\end{equation}
yielding, in combination with Eq.~\ref{eq:field_integral}, Eq.~\ref{eq:decomp} and bi-orthogonality relation, Eq.~\ref{eq:biorth}, the following system of equations for modal amplitudes $D_{m,s}$:
\begin{equation}\label{eq:single_disk_amplitude}
D_{m,s}\left(x^2-x_{m,s}^2\right)=-\frac{4\pi x^2}{n_d^2}P_{m,s}(x)-\kappa_{m,s}a_m\frac{A_{m,s}(x)}{p_m(x)}
\end{equation}
Here we introduced projections of the nonlinear polarization, $P_{m,s}$, and incident field, $A_{m,s}$, onto eigenfunctions of the disk:
\begin{eqnarray}
P_{m,s}(x)&=&\int_0^1P_m(x,\rho)\overline{\psi}_{m,s}^*(x_{m,s}n_d\rho)\rho d\rho \label{eq:Pms}\\
A_{m,s}(x)&=&\frac{2}{\pi x}\int_0^1J_m(n_dx\rho)\overline{\psi}_{m,s}^*(x_{m,s}n_d\rho)\rho d\rho\label{eq:Ams}
\end{eqnarray}
and a parameter $\kappa_{m,s}$ defined as
\begin{equation}\label{eq:kappa}
\kappa_{m,s}\equiv\alpha_m(x)\left(x^2-x_{m,s}^2\right)
\end{equation}
Parameter $\kappa_{m,s}$ in the last term of Eq.~\ref{eq:single_disk_amplitude} characterizes the efficiency of coupling of the incident radiation to the internal modes of the disk. The structure of this parameter, which we will call external coupling parameter, reflects distinction between external and internal excitation mechanisms:  \textbf{external} incident field induces response at scattering resonances, while \textbf{internal} excitation (polarization term) generates response at the eigenfrequencies. Taking into account Eq.~\ref{eq:alpha_approx} one can present  $\kappa_{m,s}$ in vicinity of $x_{m,s}$ as
\begin{equation}\label{eq:kappa_approx}
\kappa\approx-\frac{2i\gamma_{m,s}x_{m,s}(x-x_{m,s})}{x-x_{m,s}^{(r)}+i\gamma_{m,s}^{(r)}}.
\end{equation}
For resonances with low Q-factors the difference between eigenfrequencies and scattering resonances can be quite significant resulting in a strong frequency dependence of the coupling to the external field. However, for high-Q resonances, numerical calculations indicate (see also Ref.~\cite{ture06}) that ${\cal R}ex_{m,s}-x_{m,s}^{(r)}$ as well as ${\cal I}mx_{m,s}-\gamma_{m,s}^{(r)}$ are much smaller than the individual imaginary parts of these quantities. In this case $\kappa$ can be simplified to the following form
\begin{equation}\label{eq:kappa_limit}
\kappa\approx -2i\gamma_{m,s}x_{m,s},
\end{equation}
which shows that efficiency of coupling to the modes of the disk from outside is proportional to the width of these modes. This physically clear result expresses the fact that one cannot couple from outside to a completely closed system. However, to the best of our knowledge, this result has not been derived previously from "first principles" and in many cases (see for instance, Ref.~\cite{Chembo2010}) a factor representing the width of the resonances was introduced into the respective term "by hands".

\subsection{Nonlinear dynamics of multiple optically coupled disks in the frequency domain}
\subsubsection{General equations}
The case of multiple disks can be treated by applying  Eq.~\ref{eq:single_disk_amplitude} to each disk of the structure and including in the term containing the incident field contributions from the fields scattered by all other disks~\cite{Schmidt:2009}:
\begin{equation}\label{eq:subst}
a_m\rightarrow a_m^{p}+\sum_{v\ne p}\sum_n b_n^{v}t_{m-n}^{v,p}.
\end{equation}
Here indexes $p$,$v$ enumerate disks comprising the structure,  and $t_{m-n}^{v,p}=e^{(n-m)\theta_{v,p}}H_{m-n}(xR_{v,p}/R)$ describes optical coupling between the disks, where  $R_{v,p}$ and $\theta_{v,p}$ are, respectively, radial and angular polar coordinates of $v$-th disk relative to the $p$-th one. This terms arises from transformation of the field scattered by $v$-th disk to the coordinate system centered at $p$-th disk using Graf's addition theorem for Hankel functions~\cite{abramowitz_stegun}. If all the disks are arranged along a single line, which is chosen as a polar axis of the coordinate system, one has $\theta_{pv}=\pi$ for $p<v$ and $\theta_{pv}=0$ for $p>v$.

Eq.~\ref{eq:single_disk_amplitude} in the multiple disk case must be complimented by an equation relating scattering coefficients $b_n^p$ of different disks to each other. In the absence of nonlinearity this equation is again obtained by substituting Eq.~\ref{eq:subst} to the single disk Eq.~\ref{eq:b}, which gives~\cite{Schmidt:2009}
\begin{equation}
b_m^{p}=\alpha_m(x)\left[a_m^{p}+\sum\limits_n\sum\limits_{v\ne p}t_{m-n}^{v,p}b_n^{v}\right]\label{eq:b_mult}
\end{equation}
While the presence of nonlinear polarization in Eq.~\ref{eq:Maxwell} modifies Eq.~\ref{eq:b_mult}, this modification can be neglected since it is proportional to two relatively weak effects: inter-disk coupling and single-disk nonlinearity. Only when frequency shifts due to any of these effects becomes comparable with linear frequencies of the resonator, nonlinear corrections to Eq.~\ref{eq:b_mult} should be taken into account. Combining Eq.~\ref{eq:single_disk_amplitude} with Eq.~\ref{eq:subst} and Eq.~\ref{eq:b_mult} one can eliminate coefficients of the scattered field, $b_p^m$ from the equation for the modal amplitudes and obtain the closed equation for the latter:
\begin{widetext}
\begin{equation}\label{eq:D_multiple}
\left(x^2-x_{m,s}^2\right)\tilde{D}_{m,s}^p-\kappa_{m,s}(x)\sum\limits_n\sum\limits_{v\ne p}t_{m-n}^{v,p}\tilde{D}_{n,s}^{v}(x)=
-i\kappa_{m,s}(x)a_m^p-\frac{4\pi x^2p_m}{n_d^2A_{m,s}}\tilde{P}_{m,s}^p
\end{equation}
\end{widetext}
where we introduced renormalized modal amplitude $\tilde{D}_{m,s}^p$ related to the original amplitude defined in Eq.~\ref{eq:int_field} according to
\begin{equation}
\tilde{D}_{m,s}^p\equiv\frac{p_m}{A_{m,s}}D_{m,s}^p
\end{equation}
This amplitude describes the field outside of the resonator produced by an internal mode with amplitude $D_{m,s}^p$. It is important to note that the external coupling coefficient $\kappa_{m,s}$ appears in Eq.~\ref{eq:D_multiple} not only in the term containing incident field, but also in the one describing inter-disk coupling. This is a reflection of the obvious, but still not always appreciated, fact that the optical coupling between disks occurs via scattered rather than internal fields.

Equation ~\ref{eq:D_multiple} must be, of course, complimented by an expression for the nonlinear polarization in terms of modal amplitudes, $\tilde{D}_{m,s}$. Assuming nonlinearity of Kerr type, we can present nonlinear polarization term $P(\omega)$ as
\begin{equation}\label{eq:NLPol}
P(\mathbf{r},x)=\frac{cn_d^2n_2}{16\pi^2}\int \frac{dx_1}{2\pi}\frac{dx_2}{2\pi}F(\mathbf{r},\tilde{x}_{1,2})F(\mathbf{r},x_1)F^*(\mathbf{r},x_2)
\end{equation}
where $\tilde{x}_{1,2}\equiv x-x_1+x_2$ and we used approach of Ref.~\cite{BarclayOE:2005} to express Kerr nonlinear polarization in terms of standard nonlinear refractive index $n_2$. Combining Eq.\ref{eq:NLPol} with modal expansion Eq.~\ref{eq:int_field} one can finally derive the following expression for the component of the nonlinear polarization $\tilde{P}_{m,s}$ in terms of modal amplitudes $D_{m,s}(x)$:
\begin{widetext}
\begin{equation}\label{eq:PmsD}
\tilde{P}_{m,s}(x)=\sum_{\nu_1,\nu_2,\nu_3}\delta_{m-m_2+m_3,m_1}\Lambda_{\nu_1,\nu_2,\nu_3}^{m,s}\int\frac{dx_1}{2\pi}\frac{dx_2}{2\pi}D_{\nu_1}(\tilde{ x}_{1,2})D_{\nu_2}(x_1)D_{\nu_3}^*(x_2)
\end{equation}
\end{widetext}
where for shortness we combined double indexes $m_{i},s_{i}$ in single indexes $\nu_{i}$, and introduced nonlinear inter-mode coupling coefficients $\Lambda_{\nu_1,\nu_2,\nu_3}^{m,s}$ defined as
\begin{widetext}
\begin{equation}\label{eq:Lambda}
\Lambda_{\nu_1,\nu_2,\nu_3}^{m,s}= \frac{cn_d^2n_2}{16\pi^2}\int\limits_0^1d\rho\rho\psi_{\nu_1}(x_{\nu_1}n_d\rho)\psi_{\nu_2}(x_{\nu_2}n_d\rho)\psi_{\nu_3}(x_{\nu_3}n_d\rho)\overline{\psi}^*_{m,s}(x_{m,s}n_d\rho)
\end{equation}
\end{widetext}
The derived equations \ref{eq:D_multiple}~-~\ref{eq:Lambda} provide an \emph{ab initio} foundation for studying a variety of nonlinear processes in the the system of coupled disk resonators with Kerr nonlinearity. The next step in development of the theory would include Fourier transformation of the derived equations back to time-domain. This procedure cannot be carried out exactly and requires approximations of a "slow changing amplitude" type. The implementation of the latter, however, depends on the type of nonlinear processes being studied. In this paper we are interested in self-trapping dynamics in the system of coupled disks, thus in the subsequent sections of the paper we will apply Eq.~\ref{eq:D_multiple}~-~\ref{eq:Lambda} to this particular problem.
\subsubsection{Dynamic of modal amplitudes in a double-disk optical "molecule"}
In what follows we limit our consideration to the case of only two disks, which we will study in the so called resonant approximation~\cite{Schmidt:2009}. This approximation implies that we only take into account coupling between degenerate phase-matched modes of the disks. These are clockwise and counterclockwise modes characterized by azimuthal numbers of opposite signs. The phase matching  means that mode with $m=M$ in disk $1$ couples only to the mode $m=-M$ of the disk $2$ and vice versa. On formal level this is justified by inequality $t_{2M}\gg t_0$ valid for $M\gg 1$. In this approximation Eq.~\ref{eq:D_multiple} takes the following form:
\begin{equation}\label{eq:D_double}
\begin{split}
\left(x^2-x_{M}^2\right)\tilde{D}_{M}^1-\kappa_{m,s}(x)t_{2M}^{1,2}\tilde{D}_{-M}^{2}(x)&=-\frac{4\pi x^2p_M}{n_d^2A_{M}}\tilde{P}_{M}^1\\
\left(x^2-x_{M}^2\right)\tilde{D}_{M}^2-\kappa_{m,s}(x)t_{2M}^{1,2}\tilde{D}_{-M}^{1}(x)&=-\frac{4\pi x^2p_M}{n_d^2A_{M}}\tilde{P}_{M}^2\\
\left(x^2-x_{M}^2\right)\tilde{D}_{-M}^1-\kappa_{m,s}(x)t_{2M}^{1,2}\tilde{D}_{M}^{2}(x)&=-\frac{4\pi x^2p_M}{n_d^2A_{M}}\tilde{P}_{-M}^1\\
\left(x^2-x_{M}^2\right)\tilde{D}_{-M}^2-\kappa_{m,s}(x)t_{2M}^{1,2}\tilde{D}_{M}^{1}(x)&=-\frac{4\pi x^2p_M}{n_d^2A_{M}}\tilde{P}_{-M}^2
\end{split}
\end{equation}
where we abridged notations by replacing double index $m,s$ with a single index $M$ having in mind that in the resonant approximation all involved modes have the same radial index set to be equal to unity. In Eq.~\ref{eq:D_double} we also omitted the term with the incident field since we model excitations of the dynamics in this paper by initial conditions rather than by an external field.

Transformation of Eq.~\ref{eq:D_double} to the time domain is based on a slow amplitude approximation, when the nonlinear dynamics is presented as a  slow modulation of fast linear oscillations of the field:
\begin{equation}\label{eq:slow_amplitude_time}
 \tilde{D}_{M}^p(t) = \frac{1}{2}\left(S_M^p(t)e^{-ix_0t}+c.c.\right)
\end{equation}
where $S_M^p$ is a slow changing amplitude, $x_0$ is the frequency of the linear dynamics, and $c.c.$ means "complex conjugated" term. In the spectral domain this approximation has the following form
\begin{equation}\label{eq:slow_amplitude_spectr}
 \tilde{D}_{M}^p(x) = \frac{1}{2}\left[S_M^p(x)\delta(x-x_0)+\left[S_M^p(x)\right]^*\delta(x+x_0)\right]
\end{equation}
The choice of $x_0$ depends on the problem at hands. For instance, in the presence of an external incident field, $x_0$ would be set by its frequency, while in the case of initial value problem, considered in this work, the linear dynamics is determined by the poles of the Green's function, Eq.~\ref{eq:decomp}. These poles are solutions of equation $x_{m,s}(x)=x$, and it is clear from Eq.~\ref{eq:xi_r} and Eq.~\ref{eq:eigenvalues} that they coincide with the scattering resonances. Thus, in this case $x_0={\cal R}e[x_M^{(r)}]$ and Eq.~\ref{eq:kappa_limit} for the external coupling parameter becomes exact.

When using Eq.~\ref{eq:slow_amplitude_spectr} to convert system \ref{eq:D_double} to the time domain, one needs to take into account the dependence of the eigenfrequencies on the spectral parameter $x$ and expand it in the power series around $x=x_0$. In the slow amplitude approximation one only keeps term linear in $x$, which correspond to the first order time derivative in the time-domain:
$$
x_{M}(x)\approx x_{0}+\left.(x-x_0)\frac{dx_M}{dx}\right|_{x=x_0}
$$
While the $dx_M/dx$ term affects the coefficient in front of term with the first order time derivative of the modal amplitude, its numerical estimates showed that  in the case under consideration in this work it can be neglected.

Time-domain expression for the nonlinear polarization terms is obtained by Fourier-transforming Eq.~\ref{eq:PmsD} with the help of Eq.~\ref{eq:slow_amplitude_spectr} and restricting summation over modes only to those with $s=1$ and $m=\pm M$. In doing so, we also neglect all fast oscillating terms at combination frequencies. Combining the result with the Fourier transform of the linear part of Eq.\ref{eq:D_double} we obtain the final set of time-dependent equations for modal amplitudes
\begin{equation}
\begin{split}
i\frac{dS_{\pm M}^{(v)}(\tau)}{d\tau}+i\gamma_MS_{\pm M}^{(v)}(\tau)+h_MS_{\mp M}^{(p)}(\tau)=\\
-\zeta\Gamma_M S_{\pm M}^{(v)}\left[|S_{\pm M}^{(v)}|^2+2|S_{\mp M}^{(v)}|^2\right].
\end{split}
\label{eq:timedependent}
\end{equation}
In these equations we  introduced dimensionless time $\tau=tc/R$ and linear coupling
coefficient $h_M=i\gamma_M^{(r)}t_{2M}^{(12)}$(recall that the imaginary part of Hankel function entering $t_{2M}^{(12)}$ is much larger than its real part for $M\gg 1$, so that $h_M$ has very small imaginary part). Kerr nonlinearity results in self-coupling of modes and cross -
coupling between counter-propagating modes.
The strength of nonlinear interactions is characterized by material parameter $\zeta = (cn_2x_0)/(4\pi)$, and the dimensionless resonator enhancement parameter
\begin{equation}\label{eq:Gamma_M}
\Gamma_M = x_0R^2|A_M/p_M|^2\int\limits_0^1\left[\psi(x_M^{(r)}n_d\rho)\right]^2|\psi(x_M^{(r)}n_d\rho)|^2\rho d\rho.
\end{equation}

\begin{figure}
      \includegraphics[width= .8\linewidth, angle=0]{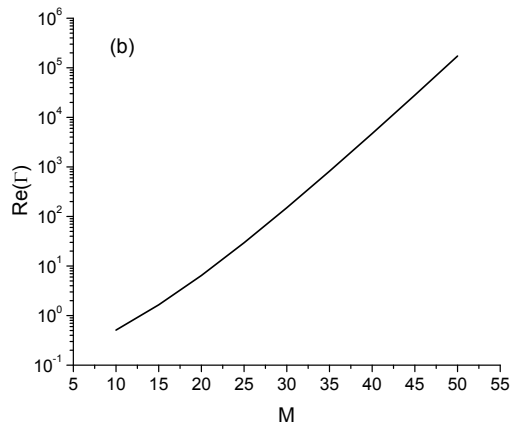}
    \caption{Nonlinear cavity enhancement factor showing significant increase over the range of $m$.
\label{fig:Gcoefficients}}
\end{figure}
\begin{figure}
   \includegraphics[width= .8\linewidth, angle=0]{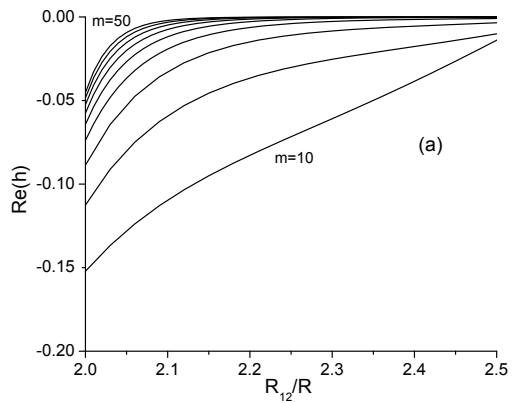}
       \caption{Linear inter-disk coupling coefficient for values of $m$ ranging from $m=10$ (lower line) to $m=50$ (upper line)
\label{fig:h_coefficient}}
\end{figure}
Both parameters $h_M$ and $\Gamma_M$ are in general complex-valued, but for high-Q modes their imaginary parts are much smaller than $\gamma_M$, which describes
the main contribution to the radiative losses of our systems. Therefore, below we assume that ${\cal I}m(h_M)={\cal I}m
(\Gamma_M)=0$. The resulting equations give a complete \emph{ab initio} description of the nonlinear dynamics of coupled
disks in the resonant approximation. A resonator-induced enhancement of nonlinear effects manifests itself through parameter $\Gamma_m$, which is found to
increase by six orders of magnitude when the azimuthal number $m$ changes from $m=5$ to $m=40$ (see Figure~\ref{fig:Gcoefficients} ). This enhancement reflects drastic concentration of the field of the WGM modes within its effective volume. Fig.~\ref{fig:h_coefficient} shows the linear inter-disk coupling parameter as a function of the inter-disk distance for values of the mode order $m$ ranging from $m=10$ to $m=50$. It is interesting, that while the nonlinear parameter shows fast growth with $m$, the linear inter-disk coupling actually decreases with $m$ while also becoming more short-ranged. This is explained by the fact that this parameter is a product of two factors: the Hankel function, which grows very fast with $m$ and the radiative decays rate $\gamma_M^{(r)}$ which significantly diminishes  with $m$. The result presented in this plot shows the interplay of these two  opposite tendencies.
\begin{figure}
   \includegraphics[width= .8\linewidth, angle=0]{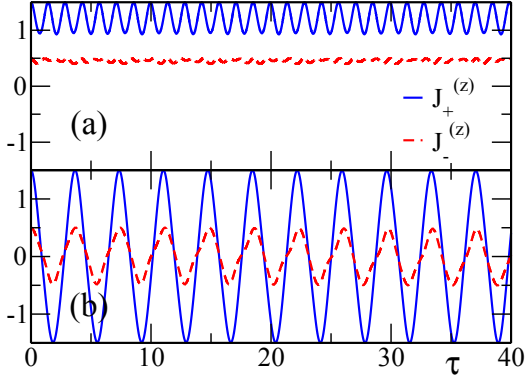}
    \caption{Imbalance intensity $J_{\pm}^{(z)}(\tau)$ for the two disks system. We have used some generic initial
conditions such that
$|S_{+}^{(1)}|^2=1.5$, $|S_{-}^{(1)}|^2=0.5$, $|S_{+}^{(2)}|^2=|S_{-}^{(2)}|^2=0$.
Plot (a)
is obtained with $\chi=2$ (self-trapping regime), while plot (b) correspond to $\chi=1$ (Josephson oscillation regime).
\label{fig:selftrapping}}
\end{figure}

\section{Self-trapping transition in the double disk system}

Non-linear discrete systems, exhibit a transition from a Josephson-like oscillatory motion to a self-trapping behavior when
the non-linearity strength increases beyond some critical value. The phenomenon has been studied extensively for a variety of
systems ranging from coupled non-linear waveguides to BEC's in optical lattices and biological systems \cite{CFK04,LSCASS08,
HT99,SGHBT10,FACTG10,SCWS08,LPWAD10,AGFHCO05,FW98,BKR03} while recently it has been also observed experimentally in the frame
of BEC's \cite{AGFHCO05}. Here, we will show that coupled optical micro-discs can be used as a prototype system to analyze
and observe experimentally such phase-transition.

Our analysis of the temporal behavior of the electric field relies on Eq.~(\ref{eq:timedependent}) where (to the first
approximation) we neglect radiative losses ($\gamma_0=0$). An analytical treatment of the dynamics can be achieved if we
re-write Eqs.~(\ref{eq:timedependent}) in terms of the Stokes parameters defined as
\begin{eqnarray}
\label{stokes}
J_{\pm}^{0}  &=&|S_{\pm 1}^{(1)}|^2 + |S_{\mp 1}^{(2)}|^2 \nonumber\\
J_{\pm}^{(x)}&=&(S_{\pm 1}^{(1)})^*S_{\mp 1}^{(2)} +S_{\pm 1}^{(1)}(S_{\mp 1}^{(2)})^*\nonumber\\
J_{\pm}^{(y)}&=& i (S_{\pm 1}^{(1)}(S_{\mp 1}^{(2)})^*-(S_{\pm 1}^{(1)})^*S_{\mp 1}^{(2)})\nonumber\\
J_{\pm}^{(z)}&=& |S_{\pm 1}^{(1)}|^2- |S_{\mp 1}^{(2)}|^2.
\end{eqnarray}
The first of these parameters, $J_{\pm}^{0}$, is associated with the total field intensity distributed between the coupled
counter-clockwise (clockwise) and clockwise (counter-clockwise) modes of discs $1$ and $2$ respectively. One can show that
$dJ_{\pm}^{(0)}/ d\tau = 0$,  i.e. $J_{\pm}^{0}$ is a conserved quantity. Of special interest is the $J_{\pm}^{(z)}$ component
which describes the intensity imbalance between counter-clockwise (clockwise) and clockwise (counter-clockwise) modes of
discs $1$ and $2$. Using the Stokes variables Eq. (\ref{stokes}), we can re-write Eqs. (\ref{eq:timedependent}) in the
following form:
\begin{equation}
\label{eq5}
\frac{d\mathbf{J}_{\pm}}{d\tau} =   \mathbf{J}_{\pm} \times \mathbf{B},
\end{equation}
where we introduced the Stokes vector $\mathbf{J}_{\pm}\equiv (J_{\pm}^{(x)},J_{\pm}^{(y)},J_{\pm}^{(z)})$, the pseudo-
magnetic field $\mathbf{B}\equiv (2,0,\chi J_{\pm}^{(z)}+2\chi J_{\mp}^{(z)})$ with $\chi=\zeta \Gamma_M/h_M$ and redefined
the time variable as $\tau\rightarrow \tau/h_M$.

For the intensity imbalance between the disks $J_{\pm}^{(z)}$, one can derive the following equation:
\begin{equation}
\label{S3evol1}
{d^2J_{\pm}^{(z)} \over d\tau^2} + 4 J_{\pm}^{(z)} -2 \chi J_{\pm}^{(x)} ( J_{\pm}^{(z)}+ 2 J_{\mp}^{(z)})=0.
\end{equation}
Direct integration of Eq. (\ref{S3evol1}) for various initial conditions indicates that there is always a critical value
$\chi_{\rm cr}$ of the non-linearity, for which $J_{\pm}^{(z)}$ shows a transition from a
regime where it is always positive or negative to a regime that oscillates around zero. An example of such behavior
is shown in Fig.~\ref{fig:selftrapping}a,b. The former case is associated to the self-trapping regime, while the latter
one is associated with the non-linear Josephson-like oscillations \cite{HT99,SGHBT10,SCWS08,LPWAD10,AGFHCO05}.
Obviously, the critical value of the
non-linearity $\chi_{\rm cr}$ above which energy transfer from one disc to another can take place, depends on the initial
preparation of the field excitation in the coupled micro-disc system.

The smallest value of $\chi_{\rm cr}$ is realized for initial conditions, $J_{\pm}^{(z)} (0)=1$ with the norm $J_{\pm}^{0}
(0)=1$. These correspond to an initial excitation of the clockwise and counterclockwise modes of one of the disks. From
Eq.~(\ref{eq5}) we find that
\begin{equation}
\label{Sxpm}
J_{+}^{(x)}+J_{-}^{(x)}=-{{\chi}\over{4}} ((J_{+}^{(z)} +J_{-}^{(z)} )^2 + 2 J_{+}^{(z)} J_{-}^{(z)})+{{3\chi}\over{2}}.
\end{equation}
Substituting the first term on the r.h.s. of Eq. (\ref{Sxpm}), from Eq. (\ref{S3evol1}), and using the fact that
$J_{+}^{(z)}$ and $J_{-}^{(z)}$ are symmetric, we eventually get
\begin{equation}
\label{duffing}
{d^2J_{+}^{(z)} \over d\tau^2} + (4 -{9\chi^2 \over 2})J_{+}^{(z)}+{9\chi^2 \over 2}{J_{+}^{(z)}}^3=0.
\end{equation}
Equation (\ref{duffing}) admits the following solution
\begin{equation}
\label{S3solution}
J_{\pm}^{(z)}(\tau)= \left\{
\begin{array}{cc}
{\rm cn} (2\tau;\eta);& \eta < 1\\
{\rm dn} (3\chi\tau/2; \eta^{-1});& \eta> 1.
\end{array}\right.
\end{equation}
where ${\rm cn}(u,\eta)$ and ${\rm dn}(u,\eta)$ are Jacobian elliptic functions, and $\eta=3\chi/4$ is the modulus of the
elliptic function. The value $\eta=1$, corresponding to $\chi_{\rm cr}=4/3$, marks a transition from an oscillatory to
a self-trapped behavior.

While the previous theoretical analysis allow us to calculate quantitatively the solutions of Eq. (\ref{eq5}) and derive
an expression for the critical non-linearity strength, we find it useful to provide also a qualitative argument explaining the existence
of Josephson to self-trapping transition.  The main observation is based
on the fact that an initial preparation will be redistributed in a way that it will minimize the Hamiltonian function
(energy) associated with our system. The latter can be derived from the equations of motion (\ref{eq:timedependent})
and has the form
\begin{eqnarray}
\label{hamil}
{\cal H}&=&\sum_{i}((S_{-}^{i-1}+S_{-}^{i+1}){S_{+}^{i}}^{*}+
(S_{+}^{i-1}+S_{+}^{i+1}){S_{-}^{i}}^{*}\nonumber \\
&&+{\chi \over 2} (|S_{+}^{i}|^4 + |S_{-}^{i}|^4)+2 \chi (|S_{+}^{i}|^2 |S_{-}^{i}|^2)),
\end{eqnarray}
In the self trapping regime, the energy is concentrated at the specifics modes which are initially populated (initial
field distribution). For such field configuration we get ${\cal H}_{\rm ST}=3\chi$. On the other hand, in the limit of
Josephson oscillations the energy is distributed (on the average) uniformly over the whole system. If one
assumes that $S_{\pm}^{(i)}\propto 1/\sqrt{2}$, we get the corresponding value of the energy function ${\cal H}_{\rm JO}
=(3\chi+4)/2$. The critical nonlinearity $\chi_{\rm cr}$ for which the transition from one regime to another occurs
can be evaluated by equating the two energy distributions. Our simple argument gives $\chi_{\rm cr}= 4/3$ which coincides
with the result obtained from the rigorous solution presented above.

This analysis, can be further extended to the case when the counter-propagating modes in each disk are allowed to interact
with each other due to, for instance, surface roughness induced scattering\cite{Gorodetsky2000,BorselliOE2005}. This
interaction can be described by adding an extra term $\nu_M S_{\mp}^{(i)}$, to the left side of Eq.(\ref{eq:timedependent}),
where $\nu_M$ describe the strength of the intra-disc coupling at each disc. Using the same analysis as above, one can
estimate the critical value of non-linearity. For example, for the initial conditions that we have used $J_{+}^{(z)}=
J_{-}^{(z)}=1$, we find $\chi_{\rm cr}=4/3$, independently from the value of the intra-disc coupling. Again, our numerical
calculations agrees with the prediction of the heuristic argument.

\section{Leaking Dynamics}

A natural question is how the self-trapping phenomenon affects the relaxation dynamics of this system once a leakage
is introduced in one of the disks.  Experimentally this can be achieved by coupling one of the disks to a tapered fiber
as discussed in the introduction. This will result in the imbalance between decay rates of the two disks so that the
smaller intrinsic decay rate in the second disk can be
neglected. Theoretically, we describe this situation by restoring a radiative decay rate, $\gamma_0$, only in the first
of the Eq. (\ref{eq:timedependent}).

\begin{figure}
   \includegraphics[width= .8\linewidth, angle=0]{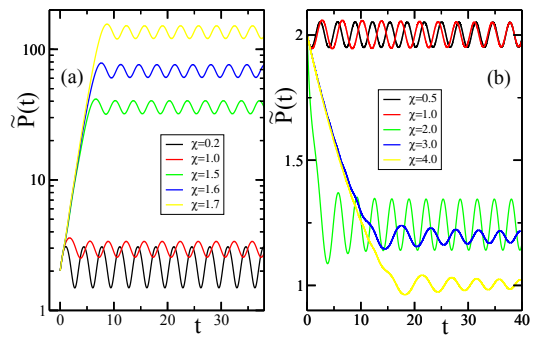}
    \caption{Temporal behaviour of the total intensity remaining inside the two discs when the first one is attached to a
fiber. Plots (a) and (b) present the rescaled total intensity ${\tilde P}(t)$ for initial excitation at the second disk
and at the first disk respectively. In (a) the leaking constant is $\gamma=0.7$ while in (b) we have used $\gamma=0.05$.
In both cases, we have used various non-linearity strengths above and below the critical non-linearity parameter.
\label{fig:relaxation1}}
\end{figure}

The object of interest is the total intensity $P(\tau)=J_{+}^{(0)}+J_{-}^{(0)}$, which is no longer a conserving quantity.
In order to eliminate effects due to trivial exponential decay of the intensity, we present the numerical results for
the relaxation dynamics in Fig.~\ref{fig:relaxation1},  in terms of the rescaled parameter  $\tilde{P} =P\exp(\gamma_0 \tau)$.
Two different relaxation regimes are observed in the short time limit. While for $\chi\leq 4/3$ the rescaled norm
oscillates with a period similar to that of the corresponding closed system in the Josephson regime, for $\chi> 4/3$
there is an exponential decay, which is not eliminated by the rescaling procedure. The origin of this discrepancy is
associated with the self-trapping phenomenon. Indeed, if initial excitation is created in the leaky disk and $\chi>4/3$,
it becomes trapped and starts leaking out at a faster rate. When, however, the total intensity decreases and the system
moves to the Josephson regime the decay process slows down since the light intensity oscillates between the leaky and the
non-leaky disk. On the other hand, if we excite the non-leaking disk, then for $\chi>4/3$, the initial decay is slower,
which manifests itself in the initial growth of the rescaled intensity ${\tilde P}$.

\begin{figure}
   \includegraphics[width= .8\linewidth, angle=0]{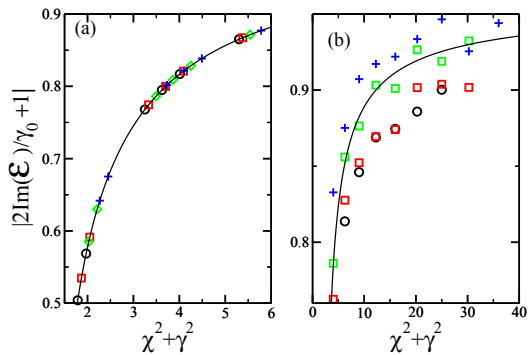}
    \caption{Decay rates extracted from the analysis of the rescaled total intensity shown in Figs. \ref{fig:relaxation1}a,b.
Plot (a) correspond to initial excitation at the second disk. Plot (b)  reports the initial decay rates when the excitation is
prepared at the first disk. In both cases, we have used various non-linearity strengths $\chi$ and leaking constants $\gamma$.
The black lines represent the theoretical predictions Eq. (\ref{reson}).
\label{fig:relaxation2}}.
\end{figure}

To achieve a better understanding of the initial decay in the self-trapped regime, we have analyzed the eigenvalue
problem associated with Eq. (\ref{eq:timedependent}) in the presence of the dissipation i.e.
\begin{eqnarray}
{\cal E}\Phi_{\pm 1}^{(i)}&=&-i\gamma_0\delta_{1,i}\Phi_{\pm 1}^{(i)}-\Phi_{\mp 1}^{(j)}\\
&&-\chi \Phi_{\pm 1}^{(i)}\left[|\Phi_{\pm 1}^{(i)}|^2+2|\Phi_{\mp 1}^{(i)}|^2\right].\nonumber
\label{eq:dissip}
\end{eqnarray}
By multiplying each set of the equations in (\ref{eq:dissip}) by the corresponding ${\Phi_{\pm 1}^{(i)}}^*$, and
adding  the resulting expressions for the propagating and counter-propagating modes associated with the two disks,
we get:
\begin{eqnarray}
\label{eq1415}
{\cal E} =
-({\Phi_{+ 1}^{(i)}}^{*} \Phi_{- 1}^{(j)} + {\Phi_{- 1}^{(j)}}^* \Phi_{+ 1}^{(i)})
- \chi (|\Phi_{+ 1}^{(i)}| ^4 +|\Phi_{- 1}^{(j)}| ^4) \nonumber \\
- 2\chi (|\Phi_{+ 1}^{(i)}|^2 |\Phi_{- 1}^{(i)}|^2
+|\Phi_{+ 1}^{(j)}|^2 |\Phi_{- 1}^{(j)}|^2) -
i\gamma_0 |\Phi_{j-i}^{(1)}|^2 \quad
\end{eqnarray}
where we have used the normalization $|\Phi_{+ 1}^{(i)}| ^2 +|\Phi_{- 1}^{(j)}| ^2 =1$ for $i\neq j$ and $i,j=1,2$. Equating
real and imaginary parts of both sides of the above equations, we get that ${\cal I}m {\cal E} = -\gamma_0 |\Phi_{+1}^{(1)}|^2=
-\gamma_0 |\Phi_{-1}^{(1)}|^2$. Using the last equality together with the normalization condition we can conclude that
$|\Phi_{-1}^{(i)}|^2=|\Phi_{+1}^{(i)}|^2$. Therefore, the four sets of equations in Eq. (\ref{eq:dissip}) reduce to the
following eigenvalue problem
\begin{eqnarray}
 {\cal E}\Phi_{+ 1}^{(1)}&=&-i\gamma_0\Phi_{+ 1}^{(1)}-\Phi_{- 1}^{(2)}
-3\chi |\Phi_{+ 1}^{(1)}|^2 \Phi_{+ 1}^{(1)}\nonumber\\
 {\cal E}\Phi_{- 1}^{(2)}&=&-\Phi_{+ 1}^{(1)}
-3\chi |\Phi_{- 1}^{(2)}|^2 \Phi_{- 1}^{(2)}
\label{eq:dissip2}
\end{eqnarray}
which can be solve exactly. Specifically, we find that for $\chi< \chi^*\equiv (1/3)\sqrt{4-\gamma_0^{2}}$ there are
two (equidistributed among the disks) leaking modes corresponding to an energy distributed equally between the two
discs $|\Phi_{+1}^{(1)}|^2=|\Phi_{-1}^{(2)}|^2$ \footnote{ We note that the difference between critical value of non-linearity for self-trapping behavior obtained from analysis presented here and that of Section III is due to the occupation distribution of the initial excitation.}. The associated complex eigenvalues ${\cal E}$ have an imaginary part
${\cal I}m ({\cal E}) = -\gamma_0/2$ which dictates their decay rate. For $\chi >\chi^{*}$ two new (non-equidistributed)
modes appear with
\begin{equation}
\label{reson}
{\cal I}m ({\cal E}^{(\pm)})=(\gamma_0/2) (-1\pm
\sqrt{1-(4/3)/[\chi^2+ \gamma_0^2]}).
\end{equation}
These solutions correspond to an non-equal energy occupations $|\Phi_{+1}^{(1)}|^2={1\over 2}(1\mp\sqrt{1-(4/3)/[
\chi^2+\gamma^2]})$. As expected, the ${\cal I}m ({\cal E}^{(+)})$ (${\cal I}m ({\cal E}^{(-)}))$ decay rate
corresponds to the mode, which has most of the intensity concentrated in the (non-)leaky disk. These rates are in
agreement with those extracted from our simulations, see Fig.~\ref{fig:relaxation2}.

\section{Conclusion}

Using \emph{ab initio} approach, we derived the equations that describe the dynamics of amplitudes
of high-Q WGMs in a system of two evanescently coupled microdisk resonators. Taking into account linear coupling only
between counter-propagating modes in adjacent disks, we studied manifestations of self-trapping phenomenon in the
relaxation dynamics. It should be noted that similar effect was also discussed in Ref.~\cite{SGHBT10,FACTG10} within the framework of quantum electrodynamics. An important distinction between the system studied in this paper and those considered in Ref.~\cite{SGHBT10,FACTG10} lies in the mechanism responsible for enhancement of nonlinear effects. In Ref.~\cite{SGHBT10,FACTG10} this enhancement was due to resonant interaction of photons with material excitations such as atoms or excitons, while in this work the enhancement occurs due to small volume of WGMs. As a result, the nonlinear effects considered in this paper can be observed at room temperature. For instance, we found that in the presence of the decay rate imbalance between the coupled disks, there exists an
anomalous relaxation behavior similar to the one discussed in  Ref.~\cite{TA96,KW04,LFO06,GKN08,NHFKG09}.  This result identifies the system of coupled optical resonators as a convenient platform for experimental study of this phenomenon, which so far, has eluded experimental observation.

\section{Acknowledgment}
(HR) and (TK) acknowledge financial support by a grant from AFOSR No. FA 9550-10-1-0433,
the DFG {\em Research Unit 760}, and by the US-Israel Binational Science Foundation (BSF), Jerusalem, Israel.
One of the authors (LD) would like to thank Arkadi Chipulin and Thomas Pertsch for their hospitality during LD's stay in Jena, where part of this work was performed.


\end{document}